\documentclass[12pt,aps,tightenlines,showpacs,nofootinbib]{revtex4}
\usepackage{amsmath,amscd,amssymb}
\usepackage[usenames,dvipsnames]{color}
\usepackage{epsfig}
\usepackage{xfrac}
\usepackage{ulem}
\usepackage{latexsym}
\usepackage{mathrsfs}
\usepackage{graphicx}
\usepackage{bbm}      % double-striked fonts for sets of integer numbers etc.
\begin{document}

\title{Wave--Packet Scattering off the Kink--Solution}

\author{A.\!\! M.\!\! H.\!\! H. Abdelhady and H. Weigel}

\affiliation{Physics Department, Stellenbosch University,
Matieland 7602, South Africa}

\begin{abstract}
We investigate the propagation of a wave--packet in the $\phi^4$ model. 
We solve the time--dependent equation of motion for two distinct initial
conditions: The wave--packet in a trivial vacuum background and in the 
background of the kink soliton solution. We extract the scattering 
matrix from the wave--packet in the kink background at very late times 
and compare it with the result from static potential scattering in the 
small amplitude approximation. We vary the size of the initial wave--packet 
to identify non--linear effects as, for example, the replacement of the 
center of the kink. 
\end{abstract}


\maketitle

\section{Introduction}

Non--linear models with soliton solutions~\cite{Ra82} possess a wide range 
of applications in physics. To present an incomplete list, this range covers 
cosmology~\cite{Vilenkin:1994,Vachaspati:2006}, particle and nuclear 
physics~\cite{Manton:2004,Weigel:2008zz}, as well as condensed matter 
physics~\cite{Bishop:1978}. The kink--model in one time and 
one space dimension for the real scalar field $\phi$ 
\begin{equation}
\mathcal{L}=\frac{1}{2}\partial_\mu \phi\, \partial^\mu \phi 
-\frac{\lambda}{4}\left(\phi^2-\frac{M^2}{2\lambda}\right)^2
\label{eq:lag}
\end{equation}
is a prototype model with a topological soliton solution based on spontaneous
symmetry breaking. Any feature observed in this model is likely to be 
relevant for the above mentioned applications.  

In the Lagrangian, eq.~(\ref{eq:lag}) $M$ is the mass parameter and $\lambda$ 
is the coupling constant. The soliton solution is the so--called \emph{kink}
\begin{equation}
\phi_{\rm K}(x)=\frac{M}{\sqrt{2\lambda}}\,
{\rm tanh}\left(\frac{M}{2}x\right)\,.
\label{eq:kink}
\end{equation}
It depends on the spatial coordinate but is time--independent and
interpolates between the degenerate vacuum configurations
$\phi_0\equiv\pm\frac{M}{\sqrt{2\lambda}}$, that arise from spontaneous symmetry 
breaking. Any deviation from either of these two values at spatial infinity 
causes the total energy 
\begin{equation}
E=\int_{-\infty}^\infty dx \, \epsilon(x,t)
\qquad{\rm with}\qquad
\epsilon(x,t)=
\frac{1}{2}\left(\dot{\phi}^2(x,t)+\phi^{\prime2}(x,t)\right)
+\frac{\lambda}{4}\left(\phi^2(x,t)-\frac{M^2}{2\lambda}\right)^2\,,
\label{eq:energy}
\end{equation}
to diverge. Here we have introduced the partial derivatives as
$\dot{\phi}=\partial \phi/\partial t$ and 
$\phi^\prime=\partial \phi/\partial x$. For later use, we have made the
time dependent energy density, $\epsilon(x,t)$ explicit. Since such divergences 
may not occur when continuously converting one configuration into another, 
any smooth deformation cannot alter the field configuration at spatial 
infinity. In particular it conserves the difference 
$\phi(-\infty,t)-\phi(\infty,t)$. This difference may thus be viewed as a 
topological charge. With proper normalization, the kink carries topological 
charge one, while the vacuum configuration $\phi(x,t)\equiv\phi_0$ has charge zero. 

Commonly small amplitude fluctuations about the kink, 
$\phi(x,t)=\phi_{\rm K}(x)+\eta(x,t)$ are introduced to determine scattering 
data. The small amplitude wave--function $\eta(x,t)$ obeys a Schr\"odinger 
type equation with a space--dependent potential induced by the kink. In this 
small amplitude approximation quadratic and higher orders in $\eta(x,t)$ 
are omitted when substituted into $\mathcal{L}$. At spatial infinity, {\it i.e.} 
far away from the kink, this Schr\"odinger equation is a Klein--Gordon equation 
with mass $M$. Application of standard methods~\cite{Ra82} to this Schr\"odinger 
problem yield the phase shift\footnote{Since the induced potential is reflectionless, 
the two eigenphase shifts are identical.}
\begin{equation}
\delta(k)=2{\rm arctan}\left(\frac{3Mk}{2k^2-M^2}\right)\,.
\label{eq:phaseshift}
\end{equation}
The generalization to three spatial dimensions is that of meson--baryon 
scattering~\cite{Schwesinger:1988af}. The kink $\phi_{\rm K}$ represents the
baryon and the fluctuations $\eta$ are mesons~\cite{Witten:1979kh}.

In this paper we show how the same result for the phase shift can be 
obtained from the physically motivated set--up of a wave--packet interacting 
with the kink. This exactly matches the scenario that a meson scatters off a 
baryon. In particular, we construct the wave--packet such that at spatial infinity
the field equals either of the possible vacuum configurations for any finite 
time $t$. In this set--up we will numerically solve the time dependent 
equation of motion
\begin{equation}
\ddot{\phi}(x,t)=\phi^{\prime\prime}(x,t)
-\lambda \left(\phi^2(x,t)-\frac{M^2}{2\lambda}\right)\phi(x,t)
\label{eq:motion}
\end{equation}
with appropriate initial and boundary conditions for $\phi(x,t)$. The 
total energy, eq.~(\ref{eq:energy}) of configurations that obey this 
equation of motion stays constant in time.

Numerical simulations of the time--dependent solutions in the kink model
have been around since quite a while. For example, the transition time of a
wave--packet through the kink was estimated in ref.~\cite{Hasenfratz:1977}.
This occurred in the greater context of modeling phase transitions. The 
resonance structure of the kink--antikink interaction was numerically studied 
in ref.~\cite{Campbell:1983xu}. Lattice studies of the kink model were reported 
in ref.~\cite{Combs:1983,Combs:1984} as a simulation of molecular dynamics. These 
numerical studies focus on kink--antikink configurations which belong to the 
topological trivial sector. More recently this sector has undergone numerical 
investigation in the context of pair--creation~\cite{Demidov:2011eu}, 
electro--weak oscillons~\cite{Adib:2002ff,Graham:2007ds} and 
bounces~\cite{Goodman:2006,Goodman:2007}.
Numerical simulations within the $2+1$ dimensional generalization at non--zero
temperature allowed to study the effect of fast quenches to model resonant 
nucleation~\cite{Gleiser:2004iy}. This scenario may have applications in 
condensed matter and cosmology~\cite{Gleiser:2006qr}. Within super--symmetric 
extensions of the kink model, numerical simulations have even been used to 
study brane world collisions~\cite{Saffin:2007qa,Saffin:2007ja,Takamizu:2006gm}.
In the unit topological charge sector numerical simulations have been employed
to investigate the effect of impurities in the kink~\cite{Kivshar:1991zz}. 
However, no study in the sector of unit topological charge is known to us 
that attempts to extract the scattering phase shift from the time--dependent 
differential equation or to identify non--linear effects in this sector by 
going beyond the small amplitude approximation for the scattering process.

This short paper is organized as follows. In the next section we describe 
the set--up for the wave--packet, {\it i.e.} we specify the initial and boundary 
conditions to solve the time dependent equation of motion~(\ref{eq:motion}). 
In principle, a single integration of this equation can, after Fourier 
transformation, provide information on scattering data for all momenta.
In section III we will discuss the numerical simulation of the time--dependent 
configuration and describe how the scattering phase shift can be extracted. 
We will observe structural changes of the interaction pattern as the amplitude
of the initial wave--packet increases. We conclude in section IV.

\section{Wave--Packet}
\label{sec2}

We will consider wave--packets in the two sectors of vanishing and of unit
topological charges. Their structures are generic and we merely have to 
specify the background fields for either of the two cases.

As initial condition for the wave--packet we consider a linear combination 
of plane waves that satisfy the Klein--Gordon dispersion relation,
$\omega_k=\sqrt{k^2+M^2}$. Hence at $t=0$ we parameterize
\begin{equation}
\eta_{\rm wp}(x)=\int_{-\infty}^\infty dk\, A(k)\, {\rm e}^{ikx}\,.
\label{eq:ftrans1}
\end{equation}
At $t=0$ the dispersion relation enters only via the
velocity of the initial wave--packet
\begin{equation}
\dot{\eta}_{\rm wp}(x)=-i\int_{-\infty}^\infty dk\,
\omega_k\,A(k)\, {\rm e}^{ikx} \,.
\label{eq:ftrans2}
\end{equation}
In the context of our numerical simulations we assume the spectral function 
in momentum space to be of Gau{\ss}ian shape
\begin{equation}
A(k)=a_0\, {\rm e}^{-\frac{(k-k_0)^2}{\sigma_k^2}}\,,
\label{eq:gauss}
\end{equation}
where $a_0$ is the amplitude, $k_0$ is the average momentum, and $\sigma_k$ is the 
width of the distribution. Though the momentum integral in eq.~(\ref{eq:ftrans1}) 
can be straightforwardly computed, we refrain from displaying it here because 
its analog in eq.~(\ref{eq:ftrans2}) has no closed representation. We compute 
either of the integrals numerically. We remark that $\eta_{\rm wp}(x)$ may 
exhibit oscillations since it is a complex Gau{\ss}ian distribution. This is 
especially the case when $\sigma_k$ is not particularly large. The distribution 
is centered around $x=0$ and $\dot{\eta}_{\rm wp}(x)$ increases with $k_0$, 
which measures the average velocity with which the wave--packet propagates.

As the equation of motion is second order in time, we have completely
determined the initial conditions for the wave--packet in eqs.~(\ref{eq:ftrans1})
and~(\ref{eq:ftrans2}). We next super--impose it with various background 
configurations that solve the static equation of motion. Such combinations serve 
as initial configurations to integrate the equation of motion~(\ref{eq:motion}).
By construction, the initial wave--packet is localized and since the total energy 
is conserved, the topological charge of this super--imposed configuration will be 
that of the background.

\subsection{Pure wave--packet}

For the pure wave--packet we merely consider the above defined 
distribution around the trivial vacuum at early times
\begin{equation}
\phi(x,0)=\frac{M}{\sqrt{2\lambda}}+\eta_{\rm wp}(x)
\qquad {\rm and} \qquad
\frac{\partial \phi(x,t)}{\partial t}\Big|_{t=0}=
\dot{\eta}_{\rm wp}(x)
\label{eq:pureWP1}
\end{equation}
and feed it into the equation of motion~(\ref{eq:motion}).
Since the velocity $\dot{\eta}_{\rm wp}(x)$ is characterized by the
dispersion relation extracted from eq.~(\ref{eq:motion}), the superposition 
\begin{equation}
\phi(x,t)=\frac{M}{\sqrt{2\lambda}}
+\int_{-\infty}^\infty dk\, A(k)\, {\rm e}^{i(kx-\omega_k t)}
\label{eq:smallamp}
\end{equation}
is an approximate solution as long as $a_0$ is small enough to neglect 
$\mathcal{O}(\eta_{\rm wp}^2)$ terms in the equation of motion. This omission
defines the small amplitude approximation. As we increase $a_0$ non--linear effects 
emerge and the solution to the differential equation will no longer be a superposition 
of plane waves.

We may converse this line of argument to study non--linear effects. Assume 
$\phi(x,t)$ to be the (numerical) solution to the equation of motion that emerges 
from the above defined initial condition. Then the deviation of the Fourier 
transform
\begin{equation}
\widetilde{\phi}_{t_f}(k)=
\frac{{\rm e}^{i\omega_k t_f}}{A(k)}\,
\int_{-\infty}^\infty\, \frac{dx}{2\pi}\, {\rm e}^{-ikx}
\left[\phi(x,t_f)-\frac{M}{\sqrt{2\lambda}}\right]
\label{eq:freeFT}
\end{equation}
from unity measures non--linear effects for times $t_f\gg0$. Obviously, a 
single integration of the equation of motion in coordinate space provides
information about the full momentum space.

\subsection{Wave--packet in kink background}

For the wave--packet with unit topological charge in the kink background 
we consider the initial configuration
\begin{equation}
\phi(x,0)=\phi_{\rm K}(x-x_0)+\eta_{\rm wp}(x)
\label{eq:solWP1}
\end{equation}
where $x_0$ is the position of the center of the kink soliton solution. It 
must be taken large enough to avoid any overlap between the kink and the 
wave--packet at $t=0$ if we want to discuss the scattering problem. In order
for scattering to occur, the signs of $x_0$ and $k_0$ must coincide. 
Otherwise the wave--packet will propagate away from the kink\footnote{The
wave--packet contains components with negative momenta. They do not
participate in scattering for our choice $x_0>0$.}. Since 
the kink is static, the initial velocity is as in eq.~(\ref{eq:ftrans2}). 
Any non--zero velocity of the kink can eventually be compensated by an 
appropriate Lorentz transformation~\cite{Ra82}, that modifies the details
but not the structure of the wave--packet.

Again, we can give an analytical expression for the solution to this initial 
condition, provided we may omit $\mathcal{O}(\eta_{\rm wp}^2)$ terms 
according to the small amplitude approximation
\begin{equation}
\phi(x,t)=\phi_{\rm K}(x-x_0)+\eta^{(S)}_{\rm wp}(x,t)\,,
\label{eq:solWP2}
\end{equation}
where
\begin{equation}
\eta^{(S)}_{\rm wp}(x,t)= \int_{-\infty}^\infty dk\, A(k)\, 
{\rm exp}\left[i\left(kx-\omega_k t+\delta(k)\right)\right]\,.
\label{eq:solWP3}
\end{equation}

As for the pure wave--packet, our analysis will be converse. 
We prescribe the spectral function 
$A(k)$ as in eq.~(\ref{eq:gauss}) together with the dispersion relation 
associated with the Klein--Gordon equation. This enters the initial 
configurations $\phi(x,0)$ and $\dot{\phi}(x,0)$ and we utilize the equation 
of motion, eq.~(\ref{eq:motion}) to find the time--dependent configuration,
$\phi(x,t)$. We then consider very late times $t_f\gg0$ at which the 
wave--packet has completely penetrated the kink and the two structures 
are again well separated and can be individually identified.  
This defines the late--time wave--packet type solution
\begin{equation}
\eta^{(S)}_{\rm wp}(x,t_f)=\phi(x,t_f)-\phi_{\rm K}(x-x_0)\,.
\label{eq:lateWP1}
\end{equation}
Its (inverse) Fourier transform should be compared with the small--amplitude 
solution, eq.~(\ref{eq:solWP3})
\begin{equation}
\int_{-\infty}^\infty\, \frac{dx}{2\pi}\, {\rm e}^{-ikx}
\eta^{(S)}_{\rm wp}(x,t_f) =
A(k)\,{\rm exp}\left[i\left(\delta(k)-\omega_k t_f\right)\right]
+\mathcal{O}(a_0^2)\,.
\label{eq:lateWP2}
\end{equation}
That is, from the numerical solution to the equation of motion we should
be able to extract the phase shift
\begin{equation}
{\rm e}^{i\delta(k)}=
\frac{{\rm e}^{i\omega_k t_f}}{A(k)}\,
\int_{-\infty}^\infty\, \frac{dx}{2\pi}\, {\rm e}^{-ikx}
\eta^{(S)}_{\rm wp}(x,t_f)\,.
\label{eq:lateWP3}
\end{equation}
As long as $\eta(x,t)$ satisfies the criteria for a small amplitude 
fluctuation, the dependence on $t_f$ cancels on the right hand side.
A main purpose of the present investigation is to compare the numerical
result, eq.~(\ref{eq:lateWP3}), for the components that participate in
scattering, with the result from small amplitude 
approximation in eq.~(\ref{eq:phaseshift}).

\section{Numerical Results}

Appropriate scaling of the coordinates and the field
\begin{equation}
(x,t)\,\longrightarrow \, \frac{(x,t)}{\sqrt{2}\,M}
\qquad {\rm and} \qquad
\phi\,\longrightarrow \, \frac{M}{\sqrt{2}\,\lambda} \,\phi
\label{eq:scale}
\end{equation}
allows us to completely absorb the model parameters. Hence their actual 
values are of minor relevance and all results are genuine. In this section
we will quote all numerical results in terms of the dimensionless 
quantities on the right hand side of eq.~(\ref{eq:scale}). In these units the 
vacuum solutions are at $\phi_0=\pm1$ and the small amplitude fluctuations
have mass~$\sqrt{2}$.

The numerical treatment starts by defining an equi--distant grid with spacing~$h$ 
in coordinate space. This establishes an interval on the $x$--axis that we take 
to be finite but much larger than the extension of the wave--packet and the kink. 
Then we employ a fourth order Runge--Kutta algorithm together with an adaptive 
step size control to solve the equation of motion~(\ref{eq:motion}). The latter
considerably slows down the numerical computation as the amplitude $a_0$ is
increased. Earlier attempts using a simple Euler algorithm failed to produce 
acceptable accuracy. The equation of motion propagates
the configuration in time. At each time step (as well as at the auxiliary 
intermediate steps required by the Runge--Kutta algorithm) we compute the 
(second) spatial derivative of the configuration that occurs on the 
right--hand--side of the equation of motion with an $\mathcal{O}(h^4)$ accuracy. 
To this end, the configuration is assumed to vanish at points outside the
considered interval in coordinate space. This corresponds to the boundary
condition that no flux penetrates outside this interval. As a consequence 
thereof, the wave--packet bounces at the spatial boundaries after very long 
times. This is, of course, not physical but merely a finite size effect and 
we have to terminate the simulation at late times when this phenomena becomes 
visible.

\begin{figure}[t]
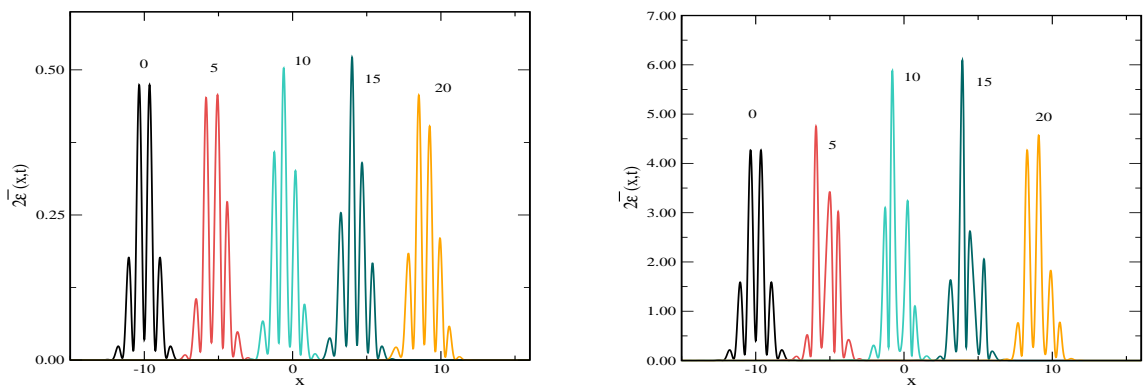

\vspace{0.6cm}
\centerline{
\epsfig{file=edwo05.eps,width=7cm,height=5cm}\hspace{1cm}
\epsfig{file=edwo15.eps,width=7cm,height=5cm}~~}
\vspace{0.2cm}
\caption{\label{fig:edensw0}(Color online)  Time snapshots of the energy
density $\overline{\epsilon}(x,t)$ (in $M^4/8\lambda^2$) of the wave--packet 
for real initial conditions.  We have used $k_0=4$ and $\sigma_k=2$. 
Left panel: $a=0.05$, right panel: $a=0.15$. The numbers next to the lines
refer to the time variable. Note the different scales on 
the ordinate.} 
\end{figure}

\begin{figure}[t]
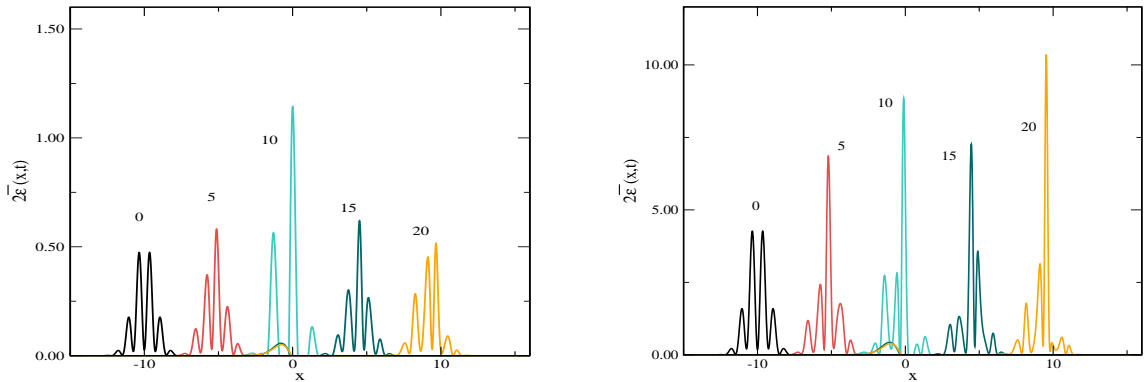

\vspace{0.6cm}
\centerline{
\epsfig{file=edwk05.eps,width=7cm,height=5cm}\hspace{1cm}
\epsfig{file=edwk15.eps,width=7cm,height=5cm}~~}
\vspace{0.1cm}
\caption{\label{fig:edenswk}(Color online) Same as figure~\ref{fig:edensw0} 
with the kink background.}
\end{figure}

A major criterion to accept the numerical solution is that the total energy, 
eq.~(\ref{eq:energy}) stays constant in time (at the order of the desired
numerical accuracy). The complex wave--packet initial condition implies 
the total energy and the energy density to be complex as well. The 
investigation of the physical energy density, $\epsilon(x,t)$ 
in eq.~(\ref{eq:energy}) hence requires to also solve the equation of 
motion with the real initial condition 
\begin{equation}
\eta_R(x)=\int_{-\infty}^\infty dk\, A(k)\, {\rm cos}(kx)
\qquad{\rm and}\qquad
\dot{\eta}_R(x)=\int_{-\infty}^\infty dk\, \omega_k\, A(k)\, {\rm sin}(kx)
\label{eq:realBC}
\end{equation}
for the wave--packet. In figures~(\ref{fig:edensw0}) and~(\ref{fig:edenswk}) we 
display the time evolution of the subtracted energy density 
\begin{equation}
\overline{\epsilon}(x,t)=\epsilon(x,t)-\epsilon_{\rm bg}(x)
\label{eq:edsub}
\end{equation}
for this initial condition. To single out the wave--packet
contribution, we have subtracted the energy density associated
with the static background. For the pure wave--packet this is 
zero but with the kink background we have 
$\epsilon_{\rm bg}(x)=\frac{1}{2}
\left[1-{\rm tanh}^2\left(\frac{x-x_0}{\sqrt{2}}\right)\right]^2$ 
in the dimensionless units of eq.~(\ref{eq:scale}). 
In the context of the small amplitude approximation the kink is
assumed to be infinitely massive and thus does not change its location 
during the interaction\footnote{In the three dimensional scenario of
meson baryon scattering this resembles the large $N_C$ picture in 
which the baryon is $\mathcal{O}(N_C)$ heavier than the meson.}.
We observe an interesting effect at the center of the kink. Once the 
wave--packet has passed by, a residual deformation of the energy density
remains. This effect persists even for very long times and becomes stable.
Also, it is the more pronounced the larger the amplitude of the 
wave--packet is. We will discuss a potential explanation for this
effect in subsection B.

The energy density associated with the wave--packet in particular suggests 
to discuss its spread  by first defining the normalized expectation values
\begin{equation}
\langle x^n \rangle = \frac{\int\, dx\, x^n\, \overline{\epsilon}(x,t)}
{\int\, dx\, \overline{\epsilon}(x,t)}\,.
\label{eq:eval}
\end{equation}
This enables the computation of the (squared) standard deviation
\begin{equation}
\sigma^2=\langle x^2 \rangle-\langle x \rangle^2
\label{eq:std}
\end{equation}
as a direct measure for the width of the wave--packet. The time--dependence
of the position of the center $\langle x \rangle$ is essentially unaffected
by the kink as the data in table~\ref{tab_ave} show. Furthermore its velocity 
agrees with what is expected for the wave--packet as
$\frac{k_0}{\sqrt{k_0^2+2}}\approx0.94$ for $k_0=4$.  We find the 
total energy stored in the wave--packet to be $0.52$ and $4.65$ for 
$a_0=0.05$ and $a_0=0.15$, respectively. The latter corresponds
to more than three time the mass of a Klein--Gordon particle. Non--linear 
effects for $\langle x \rangle$, {\it i.e.} its dependence on $a_0$, 
are only marginal.  We compare the results for $\sigma$ 
from different values of the amplitude $a_0$ in both cases, with and without 
the kink background, in table~\ref{tab_std}. Obviously the kink background 
causes a significant increase of the spread of the wave--packet. Closer 
inspection shows that this manifests itself mainly after the interaction
between the wave--packet and the kink, that is for $t>10$ while even up to
the time of the interaction no significant difference between the cases
with and without the kink is observed. We assert the strong increase of
$\sigma$ after the interaction to the emergence of the above mentioned
structure around $x\approx x_0$ rather than to a direct spread of the wave--packet. 
Indeed, the comparison between figures~\ref{fig:edensw0} and~\ref{fig:edenswk}
does not indicate a severe increase of the spread.

\begin{table}
\begin{minipage}[t]{0.4\linewidth}
\centerline{
\begin{tabular}{c|rr|rr}
&\multicolumn{2}{c|}{w/o kink}&\multicolumn{2}{c}{w/ kink}\cr
\hline
$\vspace{0.2cm}t\Bigg\backslash \vspace{-0.2cm}a_0$ & 0.05 & 0.15 & 0.05 & 0.15 \cr
\hline
0~~ &-10.00 & -10.00 & -10.00 & -10.00 \cr
5~~ & -5.28 &  -5.29 &  -5.28 &  -5.27 \cr
10~~ & -0.56 &  -0.58 &  -0.55 &  -0.55 \cr
15~~ & 4.15 & 4.13 & 4.17 & 4.17 \cr
20~~ & 8.87 & 8.83 & 8.89 & 8.90 
\end{tabular}}
\caption{\label{tab_ave}The central position $\langle x\rangle$ of the 
wave--packet as a function of time for $k_0=4$ and $\sigma_k=2$.}
\end{minipage}
\hspace{2.0cm}
\begin{minipage}[t]{0.4\linewidth}
\centerline{
\begin{tabular}{c|rr|rr}
&\multicolumn{2}{c|}{w/o kink}&\multicolumn{2}{c}{w/ kink}\cr
\hline
$\vspace{0.2cm}t\Bigg\backslash \vspace{-0.2cm}a_0$ & 0.05 & 0.15 & 0.05 & 0.15 \cr
\hline
0~~ &~0.72~&~0.72~&~0.72~&~0.73~\cr
5~~ & 0.74 & 0.79 & 0.74 & 0.79 \cr
10~~ & 0.77 & 0.95 & 0.80 & 0.97\cr
15~~ & 0.84 & 1.16 & 1.65 & 1.81\cr
20~~ & 0.91 & 1.38 & 2.25 & 2.50
\end{tabular}}
\caption{\label{tab_std}The normalized standard deviation, $\sigma$, as a 
function of time.  Parameters are as for table~\ref{tab_ave}.}
\end{minipage}
\end{table}

\subsection{Propagation of pure wave--packet}

We now return to the complex valued initial wave--packet. We first consider 
the pure wave--packet, eq.~(\ref{eq:pureWP1}). The simulation of 
eq.~(\ref{eq:freeFT}) will provide information about the numerical accuracy 
that we can expect when attempting to extract the phase shift at a later stage.

\begin{figure}[t]
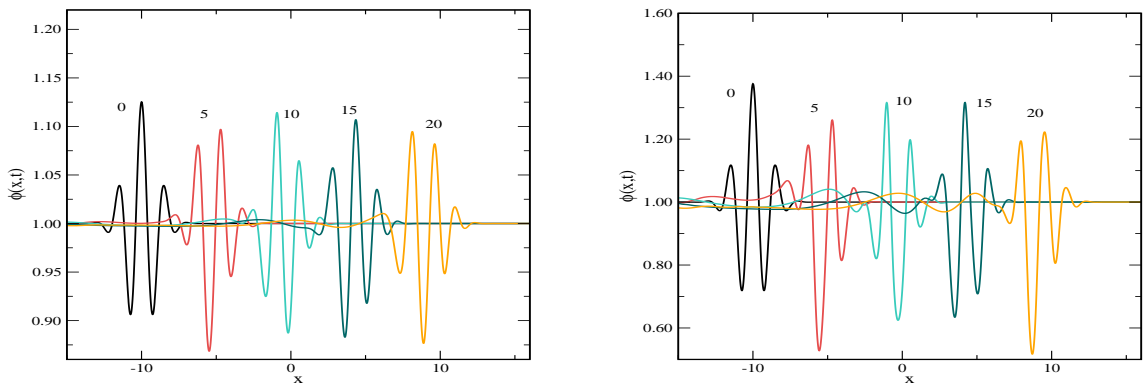

\vspace{0.6cm}
\centerline{
\epsfig{file=wo05.eps,width=7cm,height=5cm}\hspace{1cm}
\epsfig{file=wo15.eps,width=7cm,height=5cm}~~}
\vspace{0.2cm}
\caption{\label{fig:freewp}(Color online)  Time snapshots of the real 
part of the configuration $\phi(x,t)-M/\sqrt{2\lambda}$ for the initial condition, 
eq.~(\ref{eq:solWP1}).
Furthermore we have used $k_0=4$ and $\sigma_k=2$ to characterize the wave--packet. 
Left panel $a_0=0.05$, right panel $a_0=0.15$. Note the different scales 
on the ordinate.}
\vspace{0.1cm}
\end{figure}

Figure~\ref{fig:freewp} 
shows the real parts of the numerical solution to the equation of motion for two 
different values of the amplitude $a_0$. The imaginary part behaves similarly, 
just phase shifted by $\pi/2$. Note also that in addition to the expected spread, 
the number of (visible) oscillations contained within the wave--packet increases 
with time.

{}From the comparison of the two graphs in figure~\ref{fig:freewp} no 
significant dependence on the initial amplitude of the form of the 
wave--packet can be deduced. This confirms the results for the standard
deviation listed in table~\ref{tab_std}. This absence of significant 
non--linear effects is somewhat surprising. Rather they were expected 
as a signal for particle production, as sufficient energy is 
available.

\begin{figure}
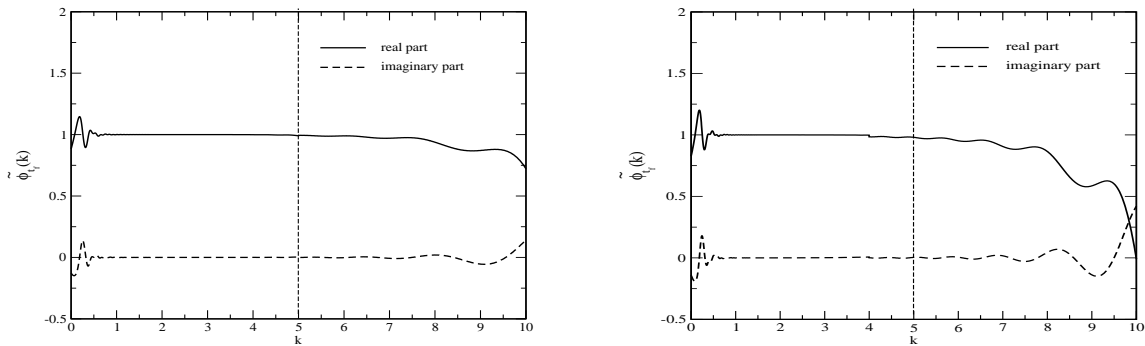

\vspace{0.4cm}
\centerline{
\epsfig{file=eq11_05.eps,width=7cm,height=4.5cm}\hspace{1cm}
\epsfig{file=eq11_20.eps,width=7cm,height=4.5cm}}
\caption{\label{fig:eq11}The inverse Fourier transform of 
eq.~(\ref{eq:freeFT}) for $a_0=0.05$ (left) and $a_0=0.20$ (right). The 
dashed vertical lines separate regimes with different numerical parameters
to improve the numerical accuracy.}
\end{figure}

In figure~\ref{fig:eq11} we show the inverse Fourier transform defined 
in eq.~(\ref{eq:freeFT}). As indicated above its deviation from unity
provides insight in the numerical accuracy that we can achieve, at least
for small $a_0$. At low momenta deficiencies arise because these contributions
have left the bulk of the wave--packet. Furthermore small errors at 
low and large momenta are amplified as $1/A(k)$ is large in these regimes. 
The deficiencies at small momenta are obviously independent of $a_0$. 
This is not the case for those at large momenta, even though their 
structures are very similar. In this momentum regime they originate from 
the inverse Fourier transform requiring a very fine grid, {\it i.e.\@} 
small $h$, in coordinate space. Unfortunately this comes with a heavy 
computational cost and it occurs appropriate to divide the momentum
axis into subintervals that are treated with different numerical 
parameters. This not only concerns the parameters for numerically solving 
the equation of motion ({\it e.g.} the step size $h$) but also the 
detailed structure of the wave--packet that we characterize by $k_0$ and 
$\sigma_k$.  At the interface of these subintervals the results extracted 
from the different numerical treatments match. This is also indicated in 
figure~\ref{fig:eq11}. At this stage it is difficult to judge whether the 
deficiencies at large~$k$ are of numerical origin or signals of the 
non--linear dynamics.

\subsection{Propagation in kink background}

\begin{figure}[t]
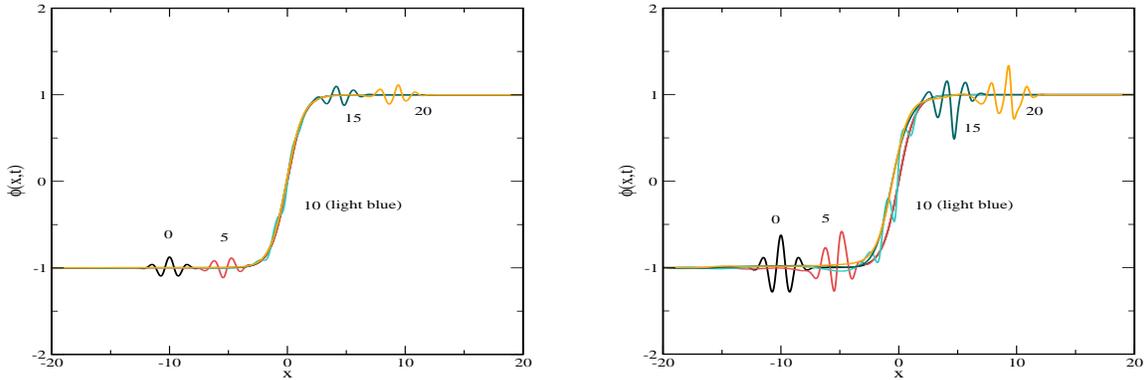

\vspace{0.6cm}
\centerline{
\epsfig{file=wk05.eps,width=7cm,height=5cm}\hspace{1cm}
\epsfig{file=wk15.eps,width=7cm,height=5cm}~~}
\vspace{0.2cm}
\caption{\label{fig:kinkwp}(Color online) Time snapshots of the real
part of the configuration $\phi(x,t)$ for the initial condition,
eq.~(\ref{eq:solWP2}). Left panel $a_0=0.05$, right panel $a_0=0.15$.}
\vspace{0.1cm}
\end{figure}

In figure~\ref{fig:kinkwp} we present solutions 
to the numerical integration of the equation of motion at different times
for small and moderate initial amplitudes of the wave--packet.
At all times, the deviation from the kink is confined within a spatial 
regime. We clearly identify the interaction process when the wave--packet 
\emph{climbs} up the kink. A wave--packet with a small amplitude essentially 
retains its shape after the interaction, the only effect being characterized 
by the phase shift that we will extract later. Surprisingly, even for moderate 
amplitudes the shape of the wave--packet does not change significantly with time. 

Figure \ref{fig:kinkwp} has the potential to explain why the energy 
density $\overline{\epsilon}$ developed some structure around $x=x_0$
after the interaction with the kink: The kink acquires
a displacement $d$. This displacement increases with the
initial amplitude~$a_0$. We quantify this increase by identifying the
point $d$ at which $\phi(d,t)=0$ in the vicinity of $x_0$ for very late 
times. We list $d$ for various values of $a_0\le0.2$ in table~\ref{tab_displace}.
In agreement and support for the small amplitude approximation, the 
displacement vanishes with the amplitude. For moderate $a_0$ it saturates
after a while and stays constant; at least at the order of our numerical
accuracy\footnote{On the overall scale this displacement is an effect 
at the order of a fraction of a percent.}.
As $a_0$ increases further, the displacement changes sign and slowly 
grows with time. This is the onset of a novel non--linear effect that we
will discuss further in subsection~\ref{sec:nl}. Our numerical simulations suggest 
that $d$ does not depend on~$k_0$.
\begin{table}
\centerline{
\begin{tabular}{c|cccc|cccc|cccc}
& \multicolumn{4}{c|}{t=100} & \multicolumn{4}{c|}{t=150} 
& \multicolumn{4}{c}{t=200} \cr 
\hline
$a_0$ & 0.05 & 0.10 & 0.15 & 0.20
& 0.05 & 0.10 & 0.15 & 0.20
& 0.05 & 0.10 & 0.15 & 0.20   \cr
\hline
$d$   & -0.07 & -0.40& -0.80 & 1.32 &
-0.07 & -0.39 & -1.13& 2.80 &
-0.07 & -0.40 & -1.09& 3.05    \cr
\end{tabular}}
\caption{\label{tab_displace}Displacement, $d$ measured relative to $x_0$, 
of the kink as a function of the initial amplitude of the wave--packet,~$a_0$.}
\end{table}

For negative $d$ the numerically observed structure (see figure~\ref{fig:edenswk})
of $\overline{\epsilon}(x,t)$ in the vicinity of $x_0$ is reasonably reproduced by 
$\epsilon_{\rm bg}(x-d)-\epsilon_{\rm bg}(x)$, {\it i.e} the corresponding shift 
of the background energy density. The larger the time interval the better is the
agreement with this analytic expression. To separate the displacement effect and 
to focus on the time--evolution of the wave--packet we repeat the calculation 
of eqs.~(\ref{eq:eval}) and~(\ref{eq:std}) with the lower boundaries of the 
integrals taken between the kink and wave--packet as we can separate these 
structures unambiguously. The results are shown in table~\ref{tab_xslate}.
\begin{table}
\centerline{
\begin{tabular}{c|cc|cc||cc|cc}
&\multicolumn{4}{c||}{w/o kink} 
&\multicolumn{4}{c}{w/ kink} \cr
\hline
&\multicolumn{2}{c|}{$\langle x\rangle$}
&\multicolumn{2}{c||}{$\sigma $} 
&\multicolumn{2}{c|}{$\langle x\rangle$}
&\multicolumn{2}{c}{$\sigma $} \cr
\hline
$\vspace{0.2cm}t\Bigg\backslash \vspace{-0.2cm}a_0$
& 0.05 & 0.15& 0.05& 0.15
& 0.05 & 0.15& 0.05& 0.15\cr
\hline
50 & 37.2 & 37.1 & 1.54 & 2.92 
   & 37.5 & 37.2 & 1.40 & 2.28 \cr
100 & 84.4 & 84.2 & 2.71 & 5.28
    & 84.8 & 85.3 & 1.84 & 4.90 \cr
150 &131.5 &131.3 & 3.89 & 7.61
    &132.1 &132.8 & 2.22 & 7.93 \cr
200 &178.7 &178.4 & 5.09 & 9.92
    &179.3 &180.2 & 4.94 &11.40 \cr
\end{tabular}}
\caption{\label{tab_xslate}Center and spread of the wave--packet in the
kink background for very late times. We compare the two cases with the 
vacuum and the kink backgrounds. In either case we have used $k_0=4$ 
and $\sigma_k=2$ with real initial conditions.}
\end{table}
The comparison with the time--evolution of the wave--packet in the
trivial background reveals that the interaction with the kink has
only a minor influence on the spread of the wave--packet once the 
displacement of the kink is properly accounted for. In both cases 
the spread increases with the amplitude of the wave--packet. This 
summarizes the main features of the wave--packet for small and 
moderate amplitudes. For $a_0>0.2$ we observe a different behavior 
that we will discuss later.

\subsection{Extraction of phase shift}

Finally we turn to a major subject of this investigation, the 
extraction of the phase shift from the scattering process. Comparison
with the result, eq.~(\ref{eq:phaseshift}), in the small amplitude 
approximation serves as a crucial test for the quality of the
numerical solution to the time--dependent equation of motion.

\begin{figure}[t]
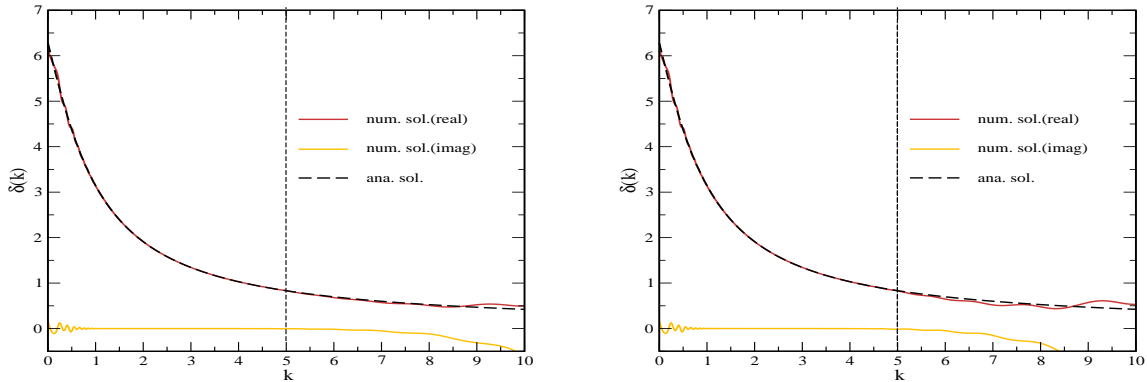

\vspace{0.6cm}
\centerline{
\epsfig{file=phase_shift_a0_0.05.eps,width=7cm,height=5cm}\hspace{1cm}
\epsfig{file=phase_shift_a0_0.15.eps,width=7cm,height=5cm}}
\vspace{0.1cm}
\caption{\label{fig:phase1}(Color online) Extraction of the phase shift 
for various momentum regimes.  Left panel: $a_0=0.05$, 
right panel: $a_0=0.15$.  Different numerical parameters
serve to improve the agreement with the analytical result in two
distinct regimes. Left regime: optimized for small and moderate momenta, 
right regime: optimized for large momenta.}
\vspace{0.1cm}
\end{figure}

Numerically solving the equation of motion~(\ref{eq:motion}) to
determine the full momentum dependence of the phase shift faces
various obstacles. First we have to incorporate the above mentioned
displacement of the kink in the integral, eq.~(\ref{eq:lateWP3}). This
is straightforwardly accomplished by restricting the integration
interval to the regime of the wave--packet. For very late times 
($t_f>100$) this regime is clearly separated from the kink. Other 
obstacles are more cumbersome. Wave--packet components with small 
momenta take a long time to finalize the interaction with the kink. 
Hence we need to solve the equation of motion for a large interval on
the time axis. In the dimensionless units defined above we consider
$t\in[0,200]$. We also vary the upper limit to ensure
stability of the results. However, components with large momenta will 
propagate a long distance in the same time interval. Hence we also 
need to consider a large interval in coordinate space. In order to 
reliably find the Fourier transform, eq.~(\ref{eq:lateWP2}) we require 
a dense grid in coordinate space for large momentum components. 
This increases the numerical cost additionally. To 
keep the numerical effort within a manageable range, it is therefore 
appropriate to split the computation in (at least) two parts. To extract 
the phase shift for small momenta, we consider a large time interval 
but a small interval in coordinate space. This leads to unreliable results
at large momenta. For that regime we consider a small time interval
but a large one in coordinate space; together with a dense grid. 
At intermediate momenta the two procedures yield identical results.
Furthermore we have the freedom to tune the parameters of the 
wave--packet, $k_0$ and $\sigma_k$ to suit the considered regime
in momentum space. These issues are indicated in figure~\ref{fig:phase1}.
In that figure we have matched two numerical treatments at $k=5$. We also 
display the numerical result for the imaginary part of the right hand 
side of eq.~(\ref{eq:lateWP2}). Its deviation from unity serves as a 
further test on the numerical accuracy. As expected this occurs for 
very small and very large momenta. Otherwise the agreement with the 
analytical result, eq.~(\ref{eq:phaseshift}) is astonishing. Certainly,
further segmentation of the momentum axis and optimization in each
segment will yield even better agreement. We have obtained the result
displayed in figure~\ref{fig:phase1} for a small amplitude ($a_0=0.05$)
for which the small amplitude approximation is expected to be accurate.
Figure~\ref{fig:phase1} shows that we find identical phase shifts 
for different amplitudes~$a_0$. Though we find differences in the 
imaginary part of the phase shift, we believe them to be short--comings
of the numerical procedure and conclude that the scattering data do not 
exhibit non--linear effects even at moderate amplitudes.

\subsection{Non--linear effects in the kink background}
\label{sec:nl}

As indicated in section~\ref{sec2} the extraction of the phase shift 
relies on the small--amplitude assumption. If $\mathcal{O}(\eta_{\rm wp}^2)$
effects are not negligible we cannot expect the right--hand--side of 
eq.~(\ref{eq:lateWP3}) to have unit magnitude. Rather we expect 
it to be less than one, corresponding to particle production. 
Yet we did not observe this effect even for moderate amplitudes, since
the wave--packet remains compact, {\it cf.} figure~\ref{fig:kinkwp}. 
For moderate amplitudes we have identified a small displacement of the kink as 
the single major effect of the non--linear dynamics. It corresponds to an 
attraction shortly before and a repulsion shortly after the interaction with 
the wave--packet.

This displacement also complicates the extraction of that part of the energy 
density, $\overline{\epsilon}(x)$ that is associated with the wave--packet. 
Once this is properly done, $\overline{\epsilon}(x)$ serves as a probability
distribution of the wave--packet. The propagation of its center does 
not exhibit consequences of the non--linear dynamics either. This propagation
is not significantly altered by the presence of the kink. However, the 
spread of the wave--function shows some increase with the amplitude of the
wave--packet.

\begin{figure}[t]
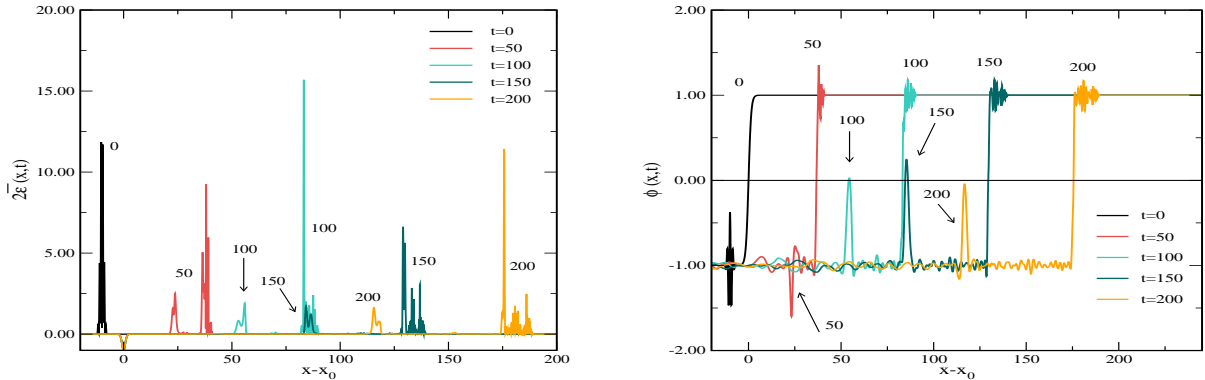

\vspace{0.6cm}
\centerline{
\epsfig{file=edcrit.eps,width=7.5cm,height=5cm}\hspace{0.8cm}
\epsfig{file=fieldcrit.eps,width=7.5cm,height=5cm}~~~}
\vspace{0.1cm}
\caption{\label{fig:edcrit}(Color online) Results for real initial conditions 
using $a_0=0.25$. The numbers next to the lines refer to the time variable.
Note that the second structure at $t=150$ is partially covered by the main
structure at $t=100$. Left panel: subtracted energy density $\overline{\epsilon}(x,t)$,
right panel: field configuration $\phi(x,t)$.}
\end{figure}

We show the typical behavior of the energy density, 
$\overline{\epsilon}(x,t)$ in figure~\ref{fig:edcrit} as we further 
increase the amplitude. Surprisingly, there is no 
footprint from the kink at $x_0$. Instead we observe that the energy density
splits into two pieces of different velocities and there is a dominant peak in 
$\overline{\epsilon}(x,t)$ at the back of the more quickly propagating piece.
The field configuration itself reveals the answer to this peculiarity.
First we observe that also the wave--packet splits in (at least) two pieces 
of different velocities. This is an indication for particle production.
Actually we also observe such splitting for the trivial background,
however, it sets in at somewhat larger amplitude\footnote{For the vacuum 
background we did not observe it for $a_0=0.25$ but for $a_0=0.35$.}. More 
notably, the transition from one vacuum configuration ($\phi_0=-1$) to the 
other ($\phi_0=1$), which is required to occur somewhere by conservation 
of the topological charge, now emerges at the back of the dominant piece 
of the wave--packet rather than at the position of the kink before the 
interaction.  Obviously the dominant peak in $\overline{\epsilon}(x,t)$ 
results from the kink being dragged by the wave--packet.
As a pronounced non--linear effect we find the transition from a 
stationary kink to a co--moving kink as the amplitude exceeds a 
certain value $a_c$. We extract this value from the behavior of the energy
density. When a structure persists at a value of $d\approx x_0$ that varies
only marginally in time the kink is considered to be stationary. Disappearance 
of this structure provides the critical value. For $k_0=4$ and $\sigma_k=2$ 
we find $a_c=0.201$. Obviously, it is impossible to extract scattering
data from these structures. This is even more the case as parts of the 
initial wave--packet now trail after the kink.

\section{Summary and Outlook}

We have performed numerical simulations in the kink--soliton model.
In particular we have set up the initial conditions to investigate the 
interaction of the kink with a wave--packet. Within this approach we
have reconstructed the phase shift known from potential scattering 
in the small amplitude approximation. By choosing a significantly wide 
initial spectral function (in momentum space) this turned possible by a 
single integration of the equation of motion in coordinate space. However, 
numerical accuracy can be improved by optimizing both, numerical 
parameters and initial conditions, to a special momentum regime.
This technique makes no particular reference to the background configuration
and can hence be employed to other cases for which no analytic result
is available. Although our technique also captures non--linear effects 
we did not observe them for the phase shift in the regime of small and 
moderate amplitude of the wave--packet. In this regime the kink does not pick 
up any energy from the wave--packet. However, the kink gets slightly displaced 
as a result of the interaction. Surprisingly, the displacement is opposite to
the propagation direction of the kink. This displacement occurs as
attraction before and repulsion after the interaction. Upon further 
increase of the amplitude the kink indeed picks up kinetic energy and 
gets dragged by the wave--packet. This is our main result regarding the 
search for effects of the non--linear dynamics. We interpret this effect 
as a signal for the existence of a critical strength of the amplitude 
beyond which the attraction between the wave--packet and the kink 
inescapable. Once the kink co--moves with the
wave--packet, the scattering process cannot be uniquely identified nor
can a phase shift be extracted. 
Furthermore the wave--packet splits into distinct pieces
for sufficiently large amplitudes. We interpret this splitting as particle
production. It is also a consequence of the non--linear dynamics because
it is equally observed without the kink background.

Eventually we want to use these numerical methods to further study the 
interaction of a kink--antikink system. The comparison with the above results 
should shed some light on whether and how crossing symmetry manifests itself 
in (topological) soliton models. This is a challenging problem because
unlike in perturbative calculations it does not simply imply the rotation
of Feynman diagrams but rather relates topologically distinct sectors that 
are not connected by smooth transformations. The numerical analysis will closely 
follow the study of refs.~\cite{Demidov:2011eu,Romanczukiewicz:2010eg}.
The kink--antikink system is known to exhibit distinct 
features~\cite{Campbell:1983xu,Dorey}, depending on whether or not the 
relative velocity exceeds the critical value $v_c=0.2598$. Below $v_c$ the 
kinetic energy is not large enough to overcome the kink--antikink attraction 
and a second interaction occurs. Above $v_c$ the kinetic energy overcomes 
the barrier that causes the attraction. It will be interesting and 
challenging to establish a connection between the critical value of the 
amplitude ($a_c$) for the kink to take up kinetic energy and the above 
mentioned critical velocity ($v_c$). The existence of such a connection 
seems possible because both scenarios reflect a critical strength for
sufficient attraction to result from the non--linear interaction within 
the model.

The kink induces two bound states, one with energy 
$\omega_1=\scriptstyle{\sqrt{3}}\,\frac{M}{2}$
and another (half--bound state) with energy $\omega_1=\sqrt{2}M$~\cite{Ra82}.
It will be interesting to see whether there are connections between
these bound states and the co--moving kink. If so, this co--moving
object would correspond to an excited state of the kink.

\acknowledgments
One of us (AMHHA) is supported in parts by the African Institute
for Mathematical Sciences (AIMS).


\begin{thebibliography}{99}

\bibitem{Ra82}
  R.~Rajaraman, \textsl{Solitons and Instantons}.
  North Holland, Amsterdam (1982).

%\cite{Vilenkin:1994}
\bibitem{Vilenkin:1994}
  A.~Vilenkin and E.P.S.~Shellard,
  \textsl{Cosmic Strings and other Topological Defects},
  Cambridge University Press, Cambridge (UK), (1994).

%\cite{Vachaspati:2006}
\bibitem{Vachaspati:2006}
  T.~Vachaspati, \textsl{Kinks and domain walls: An introduction
  to classical and quantum solitons},
  Cambridge University Press, Cambridge (UK), (2006).

%\cite{Manton:2004}
\bibitem{Manton:2004}
  N.~Manton and P.~Sutcliffe, \textsl{Topological Solitons},
  Cambridge University Press, Cambridge, (UK), (2004).

%\cite{Weigel:2008zz}
\bibitem{Weigel:2008zz}
  H.~Weigel,
  %``Chiral Soliton Models for Baryons,''
  Lect.\ Notes Phys.\  {\bf 743} (2008) 1.
  %%CITATION = LNPHA,743,1;%%

%\cite{Bishop:1978}
\bibitem{Bishop:1978}
  A.~R. Bishop and T.~Schneider (Eds.),
  \textsl{Solitons in Condensed Matter Physics},
  Springer Verlag, Berlin (1978).

%\cite{Schwesinger:1988af}
\bibitem{Schwesinger:1988af}
  B.~Schwesinger, H.~Weigel, G.~Holzwarth and A.~Hayashi,
  %``THE SKYRME SOLITON IN PION, VECTOR AND SCALAR MESON FIELDS: pi N SCATTERING
  %AND PHOTOPRODUCTION,''
  Phys.\ Rept.\  {\bf 173} (1989) 173.
  %%CITATION = PRPLC,173,173;%%

%\cite{Witten:1979kh}
\bibitem{Witten:1979kh}
  E.~Witten,
  %``Baryons in the 1/n Expansion,''
  Nucl.\ Phys.\  B {\bf 160} (1979) 57.
  %%CITATION = NUPHA,B160,57;%%

%\cite{Hasenfratz:1977}
\bibitem{Hasenfratz:1977}
  W. Hasenfratz and R. Klein,
  % ``The Interaction of a Solitary Wave Solution with phonons
  % in a One-dimensional Model fro Displacive Structural Phase Transitions,''
  Physica {\bf 89A} (1977) 191.

%\cite{Campbell:1983xu}
\bibitem{Campbell:1983xu}
  D.~K.~Campbell, J.~F.~Schonfeld and C.~A.~Wingate,
  %``Resonance Structure in Kink - Antikink Interactions in $\phi^{4}$ Theory,''
  Physica {\bf 9D} (1983) 1.
  %%CITATION = PHYSA,9D,1;%%

%cite{Combs:1983}
\bibitem{Combs:1983}
  J.~A.~Combs and S.~Yip,
  %``Single-kink dynamics in a one-dimensional atomic chain: 
  %A nonlif inear atomistic theory and numerical simulation,''
  Phys. Rev. B {\bf 28} (1983) 6873.
%cite{Combs:1984}
\bibitem{Combs:1984}
  J.~A.~Combs and S.~Yip,
  %``Molecular dynamics study of lattice kink diffusion.''
  Phys. Rev. B {\bf 29} (1984) 438.

%\cite{Demidov:2011eu}
\bibitem{Demidov:2011eu}
  S.~V.~Demidov and D.~G.~Levkov,
  %``Soliton pair creation in classical wave scattering,''
  JHEP {\bf 1106} (2011) 016
  [arXiv:1103.2133 [hep-th]].
  %%CITATION = ARXIV:1103.2133;%%

%\cite{Adib:2002ff}
\bibitem{Adib:2002ff}
  A.~B.~Adib, M.~Gleiser and C.~A.~S.~Almeida,
  %``Long lived oscillons from asymmetric bubbles: Existence and stability,''
  Phys.\ Rev.\  D {\bf 66} (2002) 085011
  [arXiv:hep-th/0203072].
  %%CITATION = PHRVA,D66,085011;%%

%\cite{Graham:2007ds}
\bibitem{Graham:2007ds}
  N.~Graham,
  %``Numerical Simulation of an Electroweak Oscillon,''
  Phys.\ Rev.\  D {\bf 76} (2007) 085017
  [arXiv:0706.4125 [hep-th]].
  %%CITATION = PHRVA,D76,085017;%%

%\cite{Goodman:2006}
  \bibitem{Goodman:2006}
  R.~H.~Goodman and R.~Haberman,
  %``Kink-Antikink Collisions in the phi^4 Equation: 
  % The n-Bounce Resonance and the Separatrix Matrix''
  Siam J. Appl. Dyn. Syst. {\bf 4} (2006)  1195.~

%\cite{Goodman:2007}
  \bibitem{Goodman:2007}
  R.~H.~Goodman and R.~Haberman,
  %``Chaotic Scattering and the n-Bounce Resonance 
  %in Solitary-Wave Interactions''
  Phys. Rev. Lett. {\bf 98} (2007) 104103.

%\cite{Gleiser:2004iy}
\bibitem{Gleiser:2004iy}
  M.~Gleiser and R.~C.~Howell,
  %``Resonant nucleation,''
  Phys.\ Rev.\ Lett.\  {\bf 94} (2005) 151601
  [arXiv:hep-ph/0409179].
  %%CITATION = PRLTA,94,151601;%%

%\cite{Gleiser:2006qr}
\bibitem{Gleiser:2006qr}
  M.~Gleiser and R.~C.~Howell,
  %``Emergence of complex spatio-temporal behavior in nonlinear field
  %theories,''
  AIP Conf.\ Proc.\  {\bf 861} (2006) 501
  [arXiv:hep-ph/0604067].
  %%CITATION = APCPC,861,501;%%

%\cite{Saffin:2007qa}
\bibitem{Saffin:2007qa}
  P.~M.~Saffin and A.~Tranberg,
  %``The Fermion spectrum in braneworld collisions,''
  JHEP {\bf 0712} (2007) 053
  [arXiv:0710.3272 [hep-th]].~
  %%CITATION = JHEPA,0712,053;%%

%\cite{Saffin:2007ja}
\bibitem{Saffin:2007ja}
  P.~M.~Saffin and A.~Tranberg,
  %``Particle transfer in braneworld collisions,''
  JHEP {\bf 0708} (2007) 072
  [arXiv:0705.3606 [hep-th]].
  %%CITATION = JHEPA,0708,072;%%

%\cite{Takamizu:2006gm}
\bibitem{Takamizu:2006gm}
  Y.~Takamizu and K.~Maeda,
  %``Collision of domain walls in asymptotically anti de Sitter spacetime,''
  Phys.\ Rev.\  D {\bf 73} (2006) 103508
  [arXiv:hep-th/0603076].
  %%CITATION = PHRVA,D73,103508;%%

 
%\cite{Kivshar:1991zz}
\bibitem{Kivshar:1991zz}
  Y.~S.~Kivshar, Z.~Fei and L.~Vazquez,
  %``Resonant soliton-impurity interactions,''
  Phys.\ Rev.\ Lett.\  {\bf 67} (1991) 1177.
  %%CITATION = PRLTA,67,1177;%%

%\cite{Romanczukiewicz:2010eg}
\bibitem{Romanczukiewicz:2010eg}
  T.~Romanczukiewicz and Y.~Shnir,
  %``Oscillon resonances and creation of kinks in particle collisions,''
  Phys.\ Rev.\ Lett.\  {\bf 105} (2010) 081601
  [arXiv:1002.4484 [hep-th]].
  %%CITATION = PRLTA,105,081601;%%

%\cite{Dorey}
\bibitem{Dorey}
  P.~E.~Dorey, Lecture at the {\sl African Institute of 
  Mathematical Sciences} (2009), unpublished.

\end{thebibliography}
\end{document}